\documentclass[aps,prx,reprint,letterpaper,showpacs]{revtex4-1}
\usepackage{graphicx}
\usepackage{bm}
\usepackage[colorlinks,urlcolor=blue,anchorcolor=blue,linkcolor=blue,citecolor=blue,breaklinks=true]{hyperref}

\newcommand{\ignore}[1]{}

\begin{document}

\title{Phase Competition and Anomalous Thermal Evolution in High-Temperature Superconductors }
\date{\today }

\begin{abstract}
The interplay of competing orders is relevant to high-temperature superconductivity known to emerge upon suppression of a parent antiferromagnetic order typically via charge doping. How such interplay evolves at low temperature---in particular at what doping level the zero-temperature quantum critical point (QCP) is located---is still elusive because it is masked by the superconducting state. The QCP had long been believed to follow a smooth extrapolation of the characteristic temperature $T^*$ for the strange normal state well above the superconducting transition temperature. However, recently the $T^*$ within the superconducting dome was reported to unexpectedly exhibit back-bending likely in the cuprate Bi$_{2}$Sr$_{2}$CaCu$_{2}$O$_{8+\delta}$. Here we show that the original and revised phase diagrams can be understood in terms of weak and moderate competitions, respectively, between superconductivity and a pseudogap state such as $d$-density-wave or spin-density-wave, based on both Ginzburg-Landau theory and the realistic $t$-$t^{\prime}$-$t^{\prime\prime}$-$J$-$V$ model for the cuprates. We further found that the calculated temperature and doping-level dependence of the quasiparticle spectral gap and Raman response qualitatively agrees with the experiments. In particular, the $T^*$ back-bending can provide a simple explanation of the observed anomalous two-step thermal evolution dominated by the superconducting gap and the pseudogap, respectively. Our results imply that the revised phase diagram is likely to take place in high-temperature superconductors.
\end{abstract}

%\pacs{74.72.-h, 71.10.Hf, 71.27.+a, 78.30.Fs}

\author{Zuo-Dong Yu$^{1}$}
\author{Yuan Zhou$^{1,2}$}
\email{zhouyuan@nju.edu.cn}
\author{Wei-Guo Yin$^{2}$}
\email{wyin@bnl.gov}
\author{Hai-Qing Lin$^{3}$}
\author{Chang-De Gong$^{4,1}$}
\affiliation{$^1$National Laboratory of Solid State Microstructure, Department of
Physics, Nanjing University, Nanjing 210093, China\\
$^{2}$ Condensed Matter Physics and Materials Science Department, Brookhaven National Laboratory, Upton, New York 11973, U.S.A.\\
$^{3}$Beijing Computational Science Research Center, Beijing 100084, China\\
$^{4}$Center for Statistical and Theoretical Condensed Matter
Physics, Zhejiang Normal University, Jinhua 321004, China}
\maketitle

\section{introduction}
\begin{figure}[btp]
%\vspace{-0.in} \hspace{-0.0in}
\center
\includegraphics[width=\columnwidth]{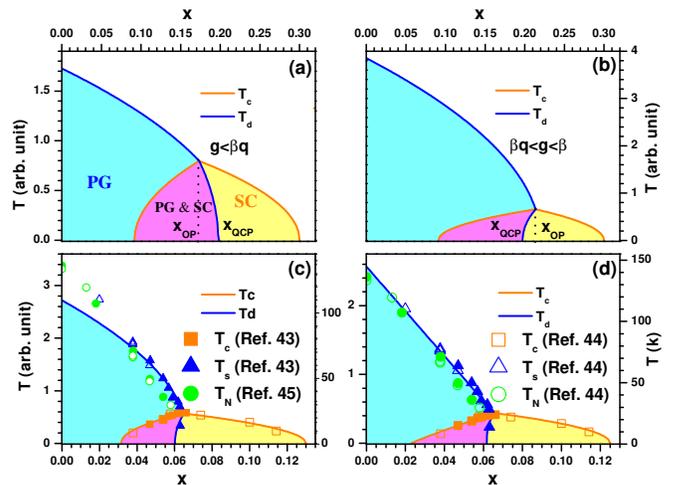}
%\vspace{-0.0in}
\caption{Phase diagrams of the competing SC (yellow), CO (cyan), and coexisting (pink) states in Ginzburg-Landau theory.  (a) The original type, where $x^{}_{\text{QCP}}>x_{\text{OP}}$, for $q=1$. (b) The revised type, where $x^{}_{\text{QCP}}<x_{\text{OP}}$, for $q=0.2$. The other parameters are $\alpha_{s}(x)=10(x-0.3)$, $\alpha_{d}(x)=27(x-0.22)$, $\beta=2$, and $g=1.2$ used in Ref.~\onlinecite{wu2005} for cuprates. Lower panels are our fits to the experimental data (various symbols) on Ba-122 iron-pnictides \cite{nandi2010,ni2008,fernandes2010} using (c) Eq.~(\ref{T2}) with $\alpha_{s}(x)=10(x-0.13)$,  $\alpha_{d}(x)=50(x-0.068)$, $q$=$0.23$, $\beta$=$2$, $g$=$1.1$, and (d) Eq.~(\ref{T}) with $\alpha_{s}(x)=10(x-0.125)$, $\alpha_{d}(x)=26(x-0.079)$, $q$=$0.4$, $p$=$0.3$, $\beta$=$2$, $g$=$1.4$.}
\label{F1}
\end{figure}

The rich phase diagrams of correlated electron materials are a central concern in both condensed matter physics and technological application \cite{dagotto05,davis13,Fradkin2015}. One archetypical example is the emerging of superconductivity (SC) upon suppression of a `parent' electronic order typically by doping. This generally yields a dome structure of the SC critical temperature $T_c$ as a function of the doping level $x$. The parent competing order (CO) ranges from the antiferormagnetic spin order in cuprates \cite{Lee2006,fischer2007,Badoux2016,Hussey2016,Rybicki2016} and heavy-fermion rare-earth compounds \cite{Jeffries2005}, to the ferro-orbital and antiferormagnetic spin dipolar/quadruplar orders in iron pnictides/chalcogenides \cite{yin_book2015,lee09,onari14,Yu2015}, and to the charge order in titanium oxypnictides \cite{frandsen14} and transition-metal dichalcogendies \cite{Morosan2006}. A particularly interesting case is the cuprate high-temperature superconductors, where the parent and SC phases do not appear to coexist but the phase competition is actually intensified by the emerging of a ``strange metal'' normal state with pseudogap opening at a temperature $T^*$ well above $T_{c}$ in the underdoped regime \cite{fischer2007}. The origin of the pseudogap has been controversial, being attributed to preformation of Cooper pairs \cite{Anderson1987,zhang1988,Kotliar1988,Emery1995,shi2008,kanigel2008} or a hidden CO such as $d$-density wave (DDW) \cite{chakravarty2001,ubbens1992,yang2006,greco2009,cappelluti1999,bejas2012}, spin-density wave (SDW) \cite{Scalapino2012,Demler2001,Moon2009,das2012}, loop-current \cite{Varma1997}, nematic or stripe order \cite{yamase2000,Kivelson2003,Fischer2011,hashimoto2010,he2011,lipscombe2009}, and pair density wave \cite{Lee2014,Berg2009}, etc. It has been observed that upon doping, $T^*$ decreases gradually in the normal state above the $T_c$ dome, and enters into the SC dome near the optimal doping level at $x^{}_{\text{OP}}$. To date, how $T^*$ evolves with doping under the $T_c$ dome is a key missing piece of the pseudogap puzzle \cite{Badoux2016,Hussey2016}. The conventional notion \cite{Lee2006,fischer2007} is that $T^*$ follows smoothly its normal-state behavior and ends ($T^*=0$) at the quantum critical point (QCP) $x^{}_{\text{QCP}} >x^{}_{\text{OP}}$ in the overdoped regime [see Fig.~\ref{F1}(a)].

However, a revised phase diagram was suggested by some recent angle-resolved photoemission spectroscopy (ARPES) measurements on Bi$_{2}$Sr$_{2}$CaCu$_{2}$O$_{8+\delta}$ (Bi-2212) cuprates \cite{vishik2012,hashimoto2015}: At slight overdoping, the system seems to change from a coexisting pseudogap-SC state to the pure SC state as temperature decreases to zero, leading to a back-bending behavior of $T^*$ as a function of $x$ under the $T_{c}$ dome [cf., Fig.~\ref{F1}(b)]. This possibility stimulates new thinking about the phase competition in the high-$T_{c}$ superconductors. For example, can the existence or nonexistence of the $T^*$ back bending be able to confirm or rule out some proposed COs as the pseudogap state? Interestingly, a similar back-bending phenomenon and revised phase diagram was clearly established in the iron-based high-$T_c$ superconductor Ba(Fe$_{1-x}$Co$_{x}$)$_{2}$As$_{2}$ (Ba-122) \cite{nandi2010,ni2008,fernandes2010}, where the QCP is located at the underdoped region, i.e., $x^{}_{\text{QCP}}<x_{\text{OP}}$  [see Figs.~\ref{F1}(c)-(d)], although undoped iron pnictides are bad metals rather than Mott insulators like cuprates.

Theoretically, a back bending of $T^*$ was obtained in a simple Landau theory for certain competition between two orders \cite{wu2005}. Thus, the revised phase diagram can happen in principle, but whether it does take place in real materials or the realistic microscopic models for them remains elusive. A mean-field-type theory of the $t$-$J$ model for the cuprates \cite{cappelluti1999,bejas2012} predicted a ``pre-back-bending'' of $T^*$ due to DDW, namely it starts well above the $T_c$ dome and even exists without SC, in disagreement with what was suggested above by the Landau theory and the ARPES data.

The ultimate detection and comprehensive understanding of the revised phase diagram demand a study of how it is related to the many unusual spectroscopy observations. For example, previous ARPES measurements showed clear evidence that the antinodal gap enhances with temperature at optimally doped Bi$_{2}$Sr$_{2}$CuO$_{6+\delta}$ (Bi-2201) \cite{kondo2011,kondo2007} and La$_{2-x}$Sr$_x$CuO$_4$ (La-214) \cite{terashima2007}. A recent study on Bi-2201 further reported that the anomalous temperature dependence of the measured gap, from slight underdoping to slight overdoping, extends to temperatures above $T_{c}$ (below $T^*$) \cite{kaminski2015}. In comparison, the gap remains nearly unchanged below $T_{c}$ in the deeply underdoped region where the pseudogap dominates, but follows the traditional BCS-like temperature dependence in the heavily overdoped region where the SC gap dominates.
Moreover, the gap evolution can be clearly detected by electronic Raman scattering (ERS) as well. By choosing the incident and scattered light polarization vectors, one can probe the gap magnitude in different regions of the Brillouin zone (BZ). In particular, the B$_{1g}$ and B$_{2g}$ channels measure the gap features of the antinodal and nodal regions, respectively \cite{devereaux2007}. The antinodal and nodal gaps, considered to be pseudogap and SC dominated, respectively, exhibit distinct doping dependence \cite{Loret2016,guyard2008-1,guyard2008,blanc2010,le_tacon2006,sacuto2011,kanigel2006}. Their temperature evolution in slightly underdoped cuprates is rather unexpected: The gap extracted from the B$_{1g}$ channel remains nearly unchanged or even increases as temperature increases toward $T_c$, rather than decreasing to zero as predicted by the standard BCS theory for $d$-wave SC. Similar enhancement in the ERS signals were discovered in lightly underdoped iron-pnictide Ba-122 \cite{chauviere2010}, further indicating a close connection between the cuprate and iron-pnictide high-$T_{c}$ superconductors.

Here, we carry out a systematical study of the phase competition between SC and a CO using both Ginzburg-Landau theory (Section II) and different mean-field theories of the extended $t$-$J$ model for the cuprates (Section III). We show that the revised and original phase diagrams in high-$T_{c}$ superconductors can be established with the moderate and weak competitions, respectively. In the latter microscopic model, the nearest-neighbor Coulomb interaction $V$ as well as the second and third nearest-neighbor hopping integrals $t^{\prime}$ and $t^{\prime\prime}$ are included to tune the competition.  We found that the back-bending of $T^*$ under the $T_{c}$ dome is quite robust against those parameter tunings but $t^{\prime}$ is necessary to prevent the pre-back-bending of $T^*$ in the absence of SC. Inclusion of the much neglected feedback effect of SC on pseudogap can push the back-bending point from optimal doping to the overdoped regime, in better agreement with the experiments  \cite{vishik2012,hashimoto2015}. In Section IV, we calculate out the ARPES and ERS spectral functions in mean-field theory of the realistic $t$-$t^{\prime}$-$t^{\prime\prime}$-$J$-$V$ model to reveal that the back-bending of $T^*$ can provide a simple explanation of the observed anomalous temperature dependence of the antinodal gap via a two-step evolution where the SC and CO dominate low- and high-temperature regions, respectively. In Section V, we consider SDW and show that it produces a less severe back-bending of $T^*$ and worse agreement with ERS than DDW. The implications of our results are discussed in Section VI and the article is summarized in Section VII.

\section{Ginzburg-Landau theory}

To evaluate the competition between SC and a CO, we start with the standard free energy \cite{chakravarty2004,wu2005}:
\begin{equation}
F=\alpha _{s}\left( x,T\right) \left\vert \psi \right\vert ^{2}+\frac{\beta
_{s}}{2}\left\vert \psi \right\vert ^{4}+\alpha _{d}\left( x,T\right) \phi
^{2}+\frac{\beta _{d}}{2}\phi ^{4}+g\left\vert \psi \right\vert ^{2}\phi ^{2},
\label{E1}
\end{equation}
where $\psi$ and $\phi$ are the order parameter for SC and the CO, respectively; $g$ is the interaction constant between them. %When the two orders are decoupled, $|\psi|^2=-\alpha_s(x,T)/\beta_s$ and $\phi^2=-\alpha_d(x,T)/\beta_d$.
Here we use the critical temperature $T_d$ for the CO to approximate $T^*$ for the pseudogap.

For simplicity, we set $\beta_s=\beta_d=\beta$ and assume that $\alpha_{s,d}(x,T)$ are the only parameters that bear the $x$ and $T$ dependence, taking the form
\begin{equation}
\alpha_{s,d}(x,T)=\alpha_{s,d}(x)+\gamma^{(1)}_{s,d}T+\gamma^{(2)}_{s,d}T^{2}.
\end{equation}
In particular, the pure quadratic $T$ dependence introduced by Wu \textsl{et al.} \cite{wu2005} to reproduce the desired form of $\alpha_{s}(x,T)\simeq 2\beta T_{c}\left( T-T_{c}\right)$ near $T_c$ reads
\begin{eqnarray}
\alpha_{s}(x,T)&=&\alpha_{s}(x)+ \beta T^2, \nonumber \\
\alpha_{d}(x,T)&=&\alpha_{d}(x)+ q \beta T^2.
\label{T2}
\end{eqnarray}
When the two orders are decoupled, $T_c=\sqrt{-\alpha_s(x)/\beta}$ and $T_d=\sqrt{-\alpha_d(x)/q\beta}$. Here the $q$ factor describes the CO's relative tolerance to thermal suppression: The smaller $q$, the more tolerant the CO than SC\cite{wu2005}. It is shown that decreasing $q$ can change the phase diagram from the original type realized for $q > g/\beta$ [Fig.~1(a)] to the revised type realized for $q < g/\beta < 1$ [Fig.~1(b)].

\begin{figure}[tbp]
%\vspace{-0.in} \hspace{-0.0in}
\center
\includegraphics[width=\columnwidth]{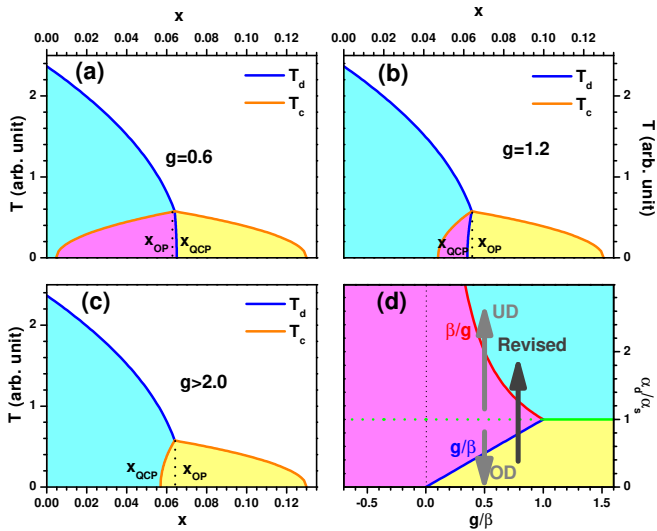}
%\vspace{-0.0in}
\caption{
Phase diagrams of the competing SC (yellow), CO (cyan), and coexisting (pink) states in Ginzburg-Landau theory using Eq.~(\ref{T2}) for fixed $q=0.4$.  (a) The original type, where $x^{}_{\text{QCP}}>x_{\text{OP}}$, for $g=0.6$. (b) The revised type, where $x^{}_{\text{QCP}}<x_{\text{OP}}$, for $g=1.2$. (c) Complete phase separation for $g>2$. The other parameters are $\alpha_{s}(x)=10(x-0.13)$, $\alpha_{d}(x)=65(x-0.069)$, $\beta=2$. (d) Phase diagram in terms of interaction $\alpha_{d}(x,T)/\alpha_{s}(x,T)$ versus $g/\beta$. The light gray, and heavy gray arrows demonstrate phase transition, which occurs in the underdoping (overdoping), and in the intermediate doping region of revised phase diagram, respectively.}
\label{F2}
\end{figure}

The phase diagram also depends sensitively on $g$, the interaction strength, as it is equally fair to read that increasing $g$ can change the phase diagram from the original type realized for $g < q\beta$ [Fig.~1(a)] to the revised type realized for $q\beta < g < \beta$ [Fig.~1(b)], providing $q<1$ and Eq.~(\ref{T2}). This is further shown in Fig.~\ref{F2} for fixed $q=0.4$. For strong enough competition ($g\geq\beta$), the two phases cannot coexist [Fig.~\ref{F2}(c)]. Therefore, the original and revised types of phase diagrams can also be generated by the weak and moderate competition between SC and other COs, respectively.

To understand the relationship between $q$ and $g$, we examine the phase diagram in terms of $\alpha_{d}(x,T)/\alpha_{s}(x,T)$ versus $g/\beta$ using Eq.~(\ref{T2}) [see Fig.~\ref{F2}(d)]. For negative $g$, the coexistence of SC and CO is the only solution, which means that the attractive interaction can generate neither the original nor the revised type of the phase diagrams found in high-$T_{c}$ superconductors. On the other hand, for strong competing interaction $g>\beta$, the two orders cannot coexist and the phase boundary is determined by $\alpha_{d}(x,T)/\alpha_{s}(x,T)=1$. For $0<g<\beta$, there are two phase boundaries in Fig.~\ref{F2}(d): The first one between the coexisting (pink) and CO (cyan) phases is set by $\alpha_{d}(x,T)/\alpha_{s}(x,T)=\beta/g>1$, and the second one between the SC (yellow) and coexisting (pink) phases is set by $\alpha_{d}(x,T)/\alpha_{s}(x,T)=g/\beta<1$. When $q=1$, the value of $\alpha_{d}(x,T)/\alpha_{s}(x,T)$ will increase as $T$ goes up if $\alpha_{d}(x,0)/\alpha_{s}(x,0)>1$, inducing the transition across the first phase boundary, as indicated by the upper gray arrow in Fig.~\ref{F2}(d). This corresponds to the underdoping scenario in the original phase diagram. Likewise when $q=1$, the value of $\alpha_{d}(x,T)/\alpha_{s}(x,T)$ will decrease as $T$ goes up if $\alpha_{d}(x,0)/\alpha_{s}(x,0)<1$, inducing the transition across the second phase boundary, as indicated by the lower gray arrow in Fig.~\ref{F2}(d). This corresponds to the overdoping scenario in the original phase diagram. To produce the revised phase diagram where the phase undergoes pure SC, coexisting, and pseudogap state as $T$ goes up, it requires that $\alpha_{d}(x,T)/\alpha_{s}(x,T)$ increases from smaller than $g/\beta$ to larger than $\beta/g$, as indicated by the black arrow in Fig.~\ref{F2}(d). Such behavior can be produced only by $q<g/\beta<1$.

We also fit the phase diagram in Ba-122 iron-pnictide though many properties of iron-based compounds differ from cuprates. However, the observed phase diagram of Ba-122 iron-pnictides \cite{nandi2010} suggests a linear $T$ dependence of $\alpha_d(x,T)$. Indeed, Eq.~(\ref{T2}) does not fit quite well [Fig.~1(c)] and a better fit [Fig.~1(d)] results from using
\begin{eqnarray}
\alpha_{s}(x,T)&=&\alpha_{s}(x)+ \frac{1-p}{\sqrt{1+p}}\sqrt{-2\alpha _{s}(x)\beta } T+\frac{2p}{1+p}\beta T^2, \nonumber \\
\alpha_{d}(x,T)&=&\alpha_{d}(x)+ q \beta T,
\label{T}
\end{eqnarray}
by which $\alpha_{s}(x,T)\simeq 2\beta T_{c}\left( T-T_{c}\right)$ near $T_c$ is still satisfied for $0\leq p < 1$. Here, $T_{c}=\sqrt{(1+p)/2}\sqrt{-\alpha _{s}(x)/\beta}$ in case of decoupling, similar to the form with the quadric $T$ dependence. This revised phase diagram also resembles the recently discovered phase diagram in Bi-2212 cuprates \cite{vishik2012,hashimoto2015}, suggesting a possible linear $T$ dependence of pseudogap in cuprate superconductors.

\section{Microscopic description in the extended $t$-$J$-$V$ model}
\begin{figure}[t]
\includegraphics[width=0.4\columnwidth]{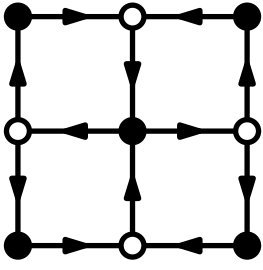}
\caption{Schematic of the DDW order. Solid, and hollow circles are for the A, and B sublattice, respectively. Signs of the DDW orders are denoted by the arrows as described in the Appendix.}
\label{A1}
\end{figure}

We proceed to study how the revised phase diagram can emerge in a microscopic theory. We focus on the $t$-$J$-type model, which was widely used to describe the low-energy physics of the cuprates \cite{Lee2006}. In particular, we examine the competition between SC and DDW/SDW. The commensurate DDW state (see Fig.~\ref{A1}) or incommensurate DDW state were shown to be the leading possible charge instability in some theories for the extended $t$-$J$-$V$ model, where $V$ is the nearest-neighbor Coulomb interaction \cite{chakravarty2001,ubbens1992,yang2006,greco2009,cappelluti1999,bejas2012,Laughlin2014a,Laughlin2014b}. $V$ is known to stabilize the DDW state with respect to phase separation \cite{cappelluti1999}.
Following the knowledge gained from the above Ginzburg-Landau theory, we also include the $V$ term to tune the robustness of the COs and the interaction strength between different orders. $V$ is chosen to reproduce the qualitative phase diagram in cuprates and its magnitude is in the same order as reported in first-principle studies of cuprates \cite{yin09}. Considerable $V$ can originate from three sources, which will be discussed later in Section VI. The extended $t$-$J$-$V$ model reads
\begin{eqnarray}
	\mathcal{H}=&-&\sum_{i,j,\sigma} t_{ij}c_{i\sigma}^\dagger c_{j\sigma}+J\sum_{\langle i,j \rangle}\Big(\vec{S}_i \cdot \vec{S}_j-\frac{1}{4}n_in_j\Big)-\mu\sum_i n_i \nonumber\\
&+&V\sum_{\langle i,j \rangle}n_{i}n_{j},
\label{E2}
\end{eqnarray}
where $c_{i\sigma}^{\dagger}$ and $c_{i\sigma}$ are electron creation and annihilation operators, respectively, at the $i$th lattice site with the constraint of single occupation. $t_{ij}$ is the hopping integral between the $i$th and $j$th sites. $J$ is the antiferromagnetic superexchange coupling constant between nearest-neighbor spins.
%$V$ is the effective interaction between nearest-neighbor charges.
We also consider the \emph{1}st, \emph{2}nd, and \emph{3}rd nearest-neighbor hopping integrals ($t$, $t^{\prime}$, and $t^{\prime\prime}$, respectively) for $t_{ij}$ to tune the shape of the Fermi surface, which is a fundamental microscopic factor underlying the phase competition.

\begin{figure}[btp]
%\vspace{-0.0in} \hspace{-0.0in}
\center
\includegraphics[width=\columnwidth]{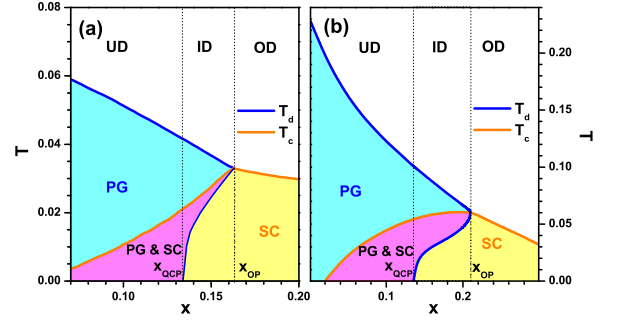}
%\vspace{-0.0in}
\caption{Phase diagram of the extended $t$-$J$-$V$ model in the color codes of yellow (SC), cyan (DDW), pink (coexisting). (a) Results in the slave-boson approximation for $t^{\prime}=-0.25$, $t^{\prime\prime}=0.1$, $J=0.35$, and $V=0.12$; (b) Results in the renormalized mean-field theory approximation for $t^{\prime}=-0.2$, $t^{\prime\prime}=0.1$, $J=0.3$, and $V=0.095$.  Besides the conventional underdoping (UD) and overdoping (OD) region, an intermediate doping (ID) region where the back-bending of $T_d$ occurs is marked out.}
\label{F3}
\end{figure}

We introduce the mean-field order parameters as $\langle c_{i}^\dagger c_{j}\rangle=\chi\pm iD$ and $\frac{1}{2}\langle c_{i\uparrow} c_{j\downarrow}-c_{i\downarrow} c_{j\uparrow}\rangle=\pm \Delta$
with $\chi$, $D$, and $\Delta$ are the uniform bond, DDW, and \emph{d}-wave SC order, respectively (see Appendix A for details). For simplicity, we adopt the slave-boson method \cite{ubbens1992}, which directly projects the original Hamiltonian into the single-occupation space via reducing the hopping terms by a factor of $x$. The order parameters can be self-consistently determined by minimizing the free energy
\begin{eqnarray}
F&=&-\frac{2T}{N}\sum_{k,\eta =\pm }^{\prime }\ln (2\cosh \frac{\beta
E_{k}^{\eta }}{2})-\mu(1-x)\nonumber\\
&+&(4V_{d}\chi ^{2}+4V_{d}D^{2}+4V_{c}\Delta ^{2})
\end{eqnarray}
with
\begin{eqnarray}
\label{V}
V_{d}&=&\frac{1}{2}J+V,\nonumber\\
V_{c}&=&J-V.
\end{eqnarray}
Here $E_{k}^{\pm}=\sqrt{\xi_{k}^{\pm 2}+\Delta_{k}^{2}}$ is the Bogliubov quasiparticle dispersion in momentum space, and  \begin{eqnarray}
\label{OP}
\Delta_{k}&=&2V_{c}\Delta (\cos k_{x}-\cos k_{y}),\nonumber \\
D_{k}&=&2V_{d}D (\cos k_{x}-\cos k_{y}).
\end{eqnarray}
$t$ has been set as the energy unit. Here we use $D_k$ to stand for the pseudogap and the DDW critical temperature $T_d$ for $T^*$.

Similar to the above macroscopic study, a revised phase diagram in hole-doped cuprates is well established within the present microscopic model. We define an intermediate doping region ranging from $x^{}_{\text{QCP}}$ to $x^{}_{\text{OP}}$ [$0.135<x<0.165$ in Fig~\ref{F3}(a)], where the back-bending of $T_d$ under the $T_{c}$ dome is found. The ground state is a pure SC state. As $T$ increases, the coexistence of the SC and DDW states emerges when the SC order parameter is sufficiently suppressed at $T_d$, which is below the $T_{c}$ dome. The magnitudes of the DDW and SC gaps are comparable in this special region. Our theoretical phase diagram qualitatively agrees with the recent laser-ARPES measurements on Bi-2212 and may also explain the discrepancy of $x^{}_{\text{QCP}}$ extrapolated by various measurements \cite{vishik2012,hashimoto2015}.

\begin{figure*}[tbp]
%\vspace{-0.0in} \hspace{-0.0in}
\centering
\includegraphics[width=1.4\columnwidth]{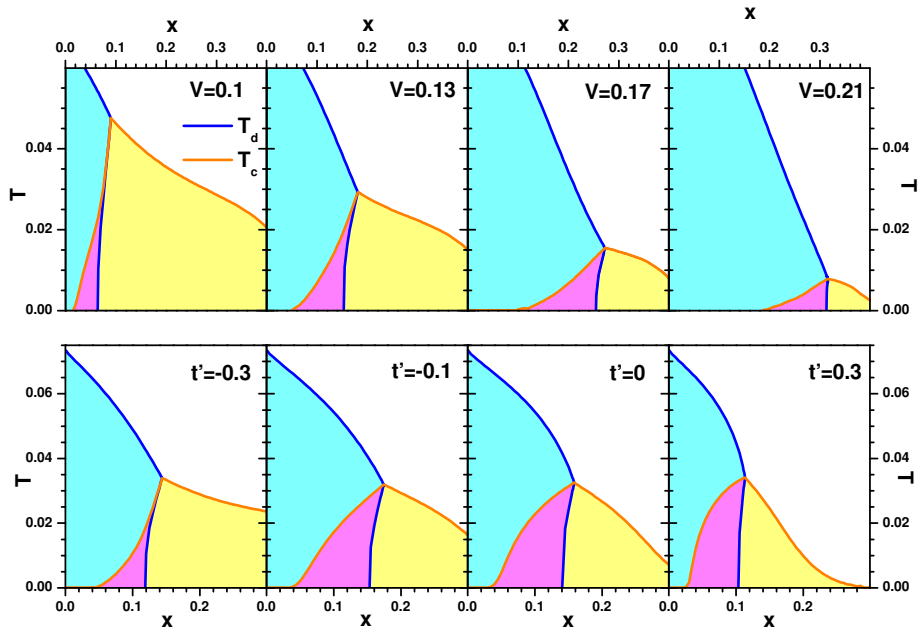}
%\vspace{-0.0in}
\caption{Model parameter dependence of the phase diagram in the color codes of yellow (SC), cyan (DDW), pink (coexisting). Top panels: The $V$ dependence for $t^{\prime}=-0.25$. Bottom panels: The $t^{\prime}$ dependence for $V=0.12$. $t^{\prime\prime}=0.1$ and $J=0.35$ for all the figures.}
\label{F4}
\end{figure*}

The $T_d$ back-bending suggests that the role of SC in the intermediate doping region has been underestimated for decades. In Fig.~\ref{F3}(b), we show the phase diagram in the renormalized mean-field approximation, which takes into account the feedback effect of SC for the renormalization of the model parameters \cite{ogata2003,wang2010} (see Appendix B). A similar back-bending phenomenon and revised diagram are obtained, indicating that the revised phase diagram is quite robust against the theoretical approximation we chose. Moreover, the DDW enters the $T_{c}$ dome now at slightly overdoping, in better agreement with experiments on Bi-2212 cuprates \cite{vishik2012,hashimoto2015}. This suggests that the feedback effect of SC be necessary to quantitative explanation of the experimental data.

Fig.~\ref{F4} shows that the presence of the $T_d$ back-bending is qualitatively robust against the variations in the model parameters, viz. $t$, $t^{\prime}$, $t^{\prime\prime}$, $J$, and $V$. According to Eq.~(\ref{V}), $V$ directly enhances the DDW order parameter and weakens SC; DDW emerges at low doping for $V>J/4$. Indeed, the calculated maximum $T_c$ decreases as $V$ increases [see the four top panels of Fig.~\ref{F4}]. $x_\mathrm{OP}-x_\mathrm{QCP}$ remains nearly unchanged for small $V$ up to 0.135 (where the maximum $T_c$ drops by half); then, it decreases as $V$ increases. This behavior is different from the $g$ or $q$ effect shown in Ginzburg-Landau theory and is attributed to the direct tuning of $\alpha_{s,d}(x,T)$ by $V$. Fig.~\ref{F4}(bottom panels) shows that the back-bending weakens as $t^{\prime}$ increases from a negative value (which means hole doping) to a positive one (which means electron doping), while the maximum $T_c$ remains nearly unchanged. Thus, the revised phase diagram could also appear in the electron-doped cuprates but it is more difficult to be detected.

\begin{figure}[!tbp]
%\vspace{-0.0in} \hspace{-0.0in}
\center
\includegraphics[width=\columnwidth]{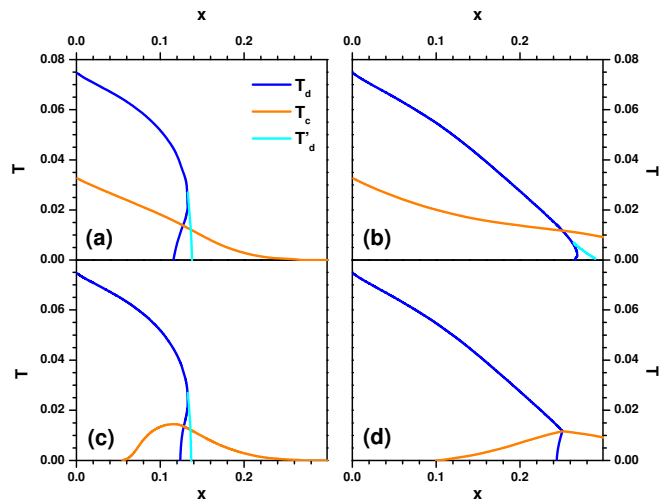}
%\vspace{-0.0in}
\caption{The effects of $t^{\prime}$ on the pre-back-bending of $T_d$ in the extended $t$-$J$-$V$ model. The DDW and SC orders are decoupled for (a) $t^{\prime}=0$ and (b) $t^{\prime}=-0.35$. They are coupled for (c) $t^{\prime}=0$ and (d) $t^{\prime}=-0.35$. T$_{d}^{\prime}$ is the characteristic temperature for the incommensurate DDW as discussed in main text. $t^{\prime\prime}=0$, $J=0.3$, and $V=0.15$ for all.}
\label{F45}
\end{figure}

A mean-field-type theory of the $t$-$J$-$V$ model with $t^{\prime}=0$ \cite{cappelluti1999,bejas2012} predicted a ``pre-back-bending'' of $T_d$ in the absence of SC. This behavior is reproduced in our calculations for $t^{\prime}=0$, as shown in Fig.~\ref{F45}(a) for the decoupled SC and DDW orders. We further found that the coupling of the SC and DDW orders suppresses the back-bending for $t^{\prime}=0$, as shown in Fig.~\ref{F45}(c). The pre-back-bending is almost entirely removed by inclusion of $t^{\prime}=-0.35$ [Fig.~\ref{F45}(b)]. In this case, the coupling of the SC and DDW orders drives the back-bending of $T_d$ [Fig.~\ref{F45}(d)].

\begin{figure}[tbp]
%\vspace{-0.0in} \hspace{-0.0in}
\center
\includegraphics[width=\columnwidth]{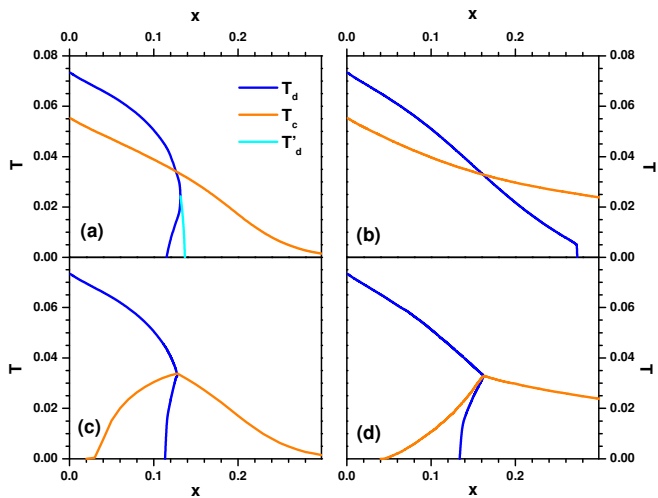}
%\vspace{-0.0in}
\caption{The effects of $t^{\prime}$ on the pre-back-bending of $T_d$ in the extended $t$-$J$-$V$ model. The DDW and SC orders are decoupled for (a) $t^{\prime}=0$ and (b) $t^{\prime}=-0.25$. They are coupled for (c) $t^{\prime}=0$ and (d) $t^{\prime}=-0.25$. T$_{d}^{\prime}$ is the characteristic temperature for the incommensurate DDW as discussed in main text. $t^{\prime\prime}=0.1$. $J=0.35$, and $V=0.12$ for all.}
\label{F46}
\end{figure}

Whether the back-bending occurs above the $T_c$ dome for $t^{\prime}=0$ \cite{cappelluti1999}, as shown in  Fig.~\ref{F45}(c), depends on the model parameters. For a smaller $V$, the back-bending starts right at $T_c$ for $t^{\prime}=0$ [see Fig.~\ref{F46}(c)], while the other features of Fig.~\ref{F45} remain unchanged in Fig.~\ref{F46}.

\emph{Incommensurate DDW}---It is previously reported that for $t=0$, the pre-back-bending of $T_{d}$ in the absence of SC vanishes upon inclusion of the incommensurate DDW, yielding a continuous decreasing of $T_{d}$ upon doping\cite{bejas2012}. We also check whether the back-bending is suppressed by the incommensurate DDW. To determine the phase boundary of the incommensurate DDW state, we study the charge instability under the random phase approximation (RPA) (see Appendix A for details).

The pre-back-bending in the normal state is removed when the incommensurate DDW is further considered as shown in Fig.~\ref{F45}(a) where only the nearest-neighbor hopping is considered, consisting with the previous results obtained by large-$N$ expansion method \cite{bejas2012}. Such an incommensurate DDW state remains for weak SC (Fig.~\ref{F45}(c)). However, the incommensurate DDW is strongly suppressed by the next nearest-neighbor hopping $t^{\prime}$ as shown in Fig.~\ref{F45}(b). Furthermore, the incommensurate DDW state is also suppressed by strong SC (Fig.~\ref{F45}(d) and Fig.~\ref{F46}(c)). Especially, the incommensurate DDW states is fully suppressed for the realistic parameters (Fig.~\ref{F46}(b) and (d)). Therefore, the back-bending of $T_{d}$ under $T_{c}$ dome presented here is driven by the interplay of SC and commensurate DDW. However, the back-bending phenomenology is parameter dependent, which may be the reason why its manifestation is found only in limited cuprates.

\section{Anomalous thermal evolution of electronic spectral features}

To explore whether and how the revised phase diagram is related to the observed anomalous temperature dependence of the antinodal gap and Raman response, we calculate these quantities in the microscopic theory.

\subsection{The quasiparticle spectral functions}
\begin{figure}[tbp]
%\vspace{-0.0in} \hspace{-0.0in}
\center
\includegraphics[width=\columnwidth]{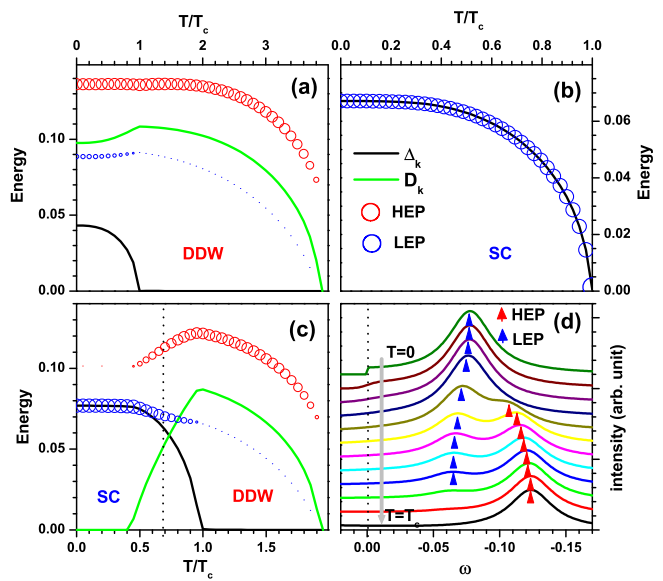}
%\vspace{-0.0in}
\caption{Thermal evolution of the SC and DDW order parameters and measured antinodal gap for three distinct doping levels: (a) underdoping $x=0.11$, (b) overdoping $x=0.18$, and (c) intermediate doping $x=0.135$. The legends of SC and DDW indicate the SC- and DDW-dominated regions, respectively. Solid lines are for the SC (black) and DDW (green) order parameters at the Fermi surface along the antinodal line. The open circles are for the peak energies extracted from the spectral functions; HEP and LEP stand for the high- and low-energy peaks, respectively. The size of the circles scale with the peak intensity. (d) Temperature evolution of the spectral functions at the intermediate doping $x=0.135$. The peak positions are marked by triangles with red for HEP and blue for LEP. The model parameters are $t^{\prime}=-0.25$, $t^{\prime\prime}=0.1$, $J=0.35$, and $V=0.12$. The Fermi energy is fixed at $0$.}
\label{F5}
\end{figure}
First, we focus on the SC and DDW order parameters $\Delta_k$ and $D_k$ [see Eq.~(\ref{OP})] and the quasiparticle spectral functions, which are the observable in ARPES measurements. Fig.~\ref{F5} shows the results at $\mathbf{k}=(k_f,0)$, the normal-state Fermi-surface momentum along the antinodal line, for three typical doping levels. For underdoping $x<0.13$ [Fig.~\ref{F5}(a)], the magnitude of the ``pseudogap'' $D_k$ is much larger than that of the SC gap $\Delta_k$. $\Delta_k$ decreases but $D_{k}$ increases as temperature increases for $T<T_{c}$. On the other hand, the gaps evaluated from the spectral functions (see Appendix) differ from the two order parameters. There exist two peaks with different weight factors below the Fermi level [c.f. Fig.~\ref{F5}(d)]; the one with substantially stronger intensity used to represent the measured gap. In the underdoped region, the high-energy peak (HEP) has much stronger intensity than the low-energy peak (LEP) and remains nearly unchanged below $T_{c}$. This reflects the fact that pseudogap dominates the underdoping region. In the overdoping region, the temperature dependence of gap follows the traditional BCS behavior since the pseudogap is absent [Fig.~\ref{F5}(b)]. These findings agree with our common knowledge and various ARPES measurements \cite{kaminski2015}.

On the contrary, in the intermediate doping range [Fig.~\ref{F5}(c)], the ``pseudogap'' $D_k$ does not emerge unless the SC gap $\Delta_k$ is suppressed sufficiently at $T_d$, similar to the previous theoretical suggestions \cite{das2008,sau2014,gabovich2014}.  On the other hand, the measured gap exhibits a pronounced two-step evolution. It evolves from the SC dominating at low temperature to the DDW dominating at high temperature [Figs.~\ref{F5}(c) and \ref{F5}(d)]. The most important feature is that the measured gap exhibits clear enhancement as temperature increases above $T_d$ (under the $T_c$ dome), especially for slight underdoping. Therefore, we find a special temperature region in the intermediate doping region where the measured gap shows anomalous temperature dependence, in good agreement with ARPES measurements on various families of cuprates \cite{kondo2011,kaminski2015,kondo2007,terashima2007}. The present explanation also differs from the previous illustrations that attribute the anomalous temperature dependence of the measured antinodal gap to either the Fermi function \cite{kordyuk2015} or the weakened SC gap \cite{yildirim2011}. Our results show that the SC gap near the borderline between the SC- and DDW-dominated regions [dotted line in Fig.~\ref{F5}(c)] only slightly weakens, in agreement with the ARPES measurements on near optimally doped Bi-2212 \cite{hashimoto2015}.

We noted  that the measured gap remains increasing even above $T_{c}$ as revealed by ARPES data \cite{kaminski2015}. This may be due to the pre-pairing of superconductivity. Although the superconducting gap and pseudogap come from different origin, the electrons may have been paired above $T_{c}$ as indicated by the ARPES \cite{kondo2011} and other experimental measurements \cite{wang2001,li2010,tallon2011}. Therefore, the back-bending phenomenon, and the region of intermediate doping is expected to be more pronounced due to strong superconducting gap magnitude.

\subsection{The Raman response}
Furthermore, we study the relationship between the revised phase diagram and the anomalous temperature dependence of ERS in the cuprates. The Raman response was calculated from using the density-density correlation function (see Appendix A). The B$_{1g}$ and B$_{2g}$ channels are contributed mainly from the Fermi surface around the antinodal and nodal regions, respectively \cite{guyard2008-1,guyard2008}. The peak energy corresponding to the B$_{2g}$ response was found to track the temperature evolution of the $d$-wave SC order due to the absence of pseudogap near the nodal region.
\begin{figure}[tbp]
%\vspace{-0.0in} \hspace{-0.0in}
\center
\includegraphics[width=\columnwidth]{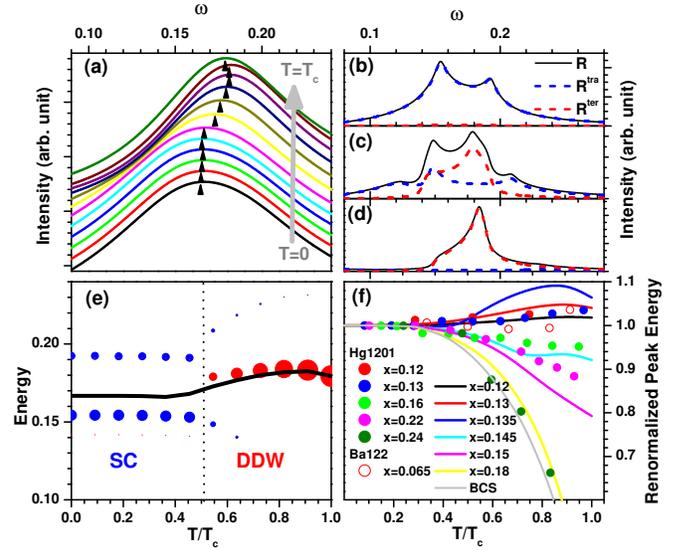}
%\vspace{-0.0in}
\caption{Thermal evolution of the Raman response in the B$_{1g}$ channel at intermediate doping $x=0.135$. (a) The results for the broadened resolution of $\eta=0.04$. The peak positions are indicated by the triangles. (b)-(d) is Raman response at three typical temperature with high resolution $\eta=0.003$. $T=0.3T_{c}$ for (b), $T=0.7T_{c}$ for (c), and $T=0.9T_{c}$ for (d). The intra-band, and inter-band components $R^{tra}$, and $R^{ter}$ is denoted by blue, and red dashed line, respectively. $R$ is the sum of $R^{tra}$ and $R^{ter}$ denoted by solid black line. (e) Temperature dependence of the energy of B$_{1g}$ Raman response peak. The circles track the peak energy shown b)-d) with high resolution, the intensity is marked by size. The solid line is the peak energy with broadened resolution extracted from (a). SC and DDW denote the SC- and DDW-dominated regions, respectively, as described in text. (f) Theoretical results under broadened resolution at different doping, together with the experimental Raman data in cuprate Hg-1201 \cite{guyard2008-1,guyard2008} and near optimally doped iron-pnictide Ba-122 \cite{chauviere2010}. $t^{\prime}=-0.25$, $t^{\prime\prime}=0.1$, $J=0.35$, and $V=0.12$.}
\label{F6}
\end{figure}

On the other hand, the Raman response in the B$_{1g}$ channel is much more complicated. In the underdoped region, the peak energy in the Raman response remains nearly unchanged with increasing temperature. It decreases monotonically with temperature and goes to zero at $T_{c}$ in the overdoping region, following a simple BCS-like temperature evolution. On the contrary, the peak energy of the Raman response in the intermediate doping region [Fig.~\ref{F6}(a)] clearly enhances upon increasing temperature toward $T_{c}$. These behaviors are qualitatively consistent with our calculated temperature dependence of the measured quasiparticle gap and the ERS measurements\cite{guyard2008-1,guyard2008,blanc2010}, where a slight upward shift of the antinodal gap component was detected in the slightly underdoped Hg1201 and Bi2212 as $T_{c}$ is approached. The discrepancy in the temperature evolution of the B$_{1g}$ and B$_{2g}$ ERS would favor the two-gap scenario.

The above single peak was obtained from using the broadened resolution of $\eta=0.04$. It is resolved into multi-peaks with $\eta=0.003$ owing to the intra-band (blue) and inter-band (red) contributions [Figs.~\ref{F6}(b)-(d)]. At low temperature ($T\ll T_{c}$) where SC dominates [Fig.~\ref{F6}(b)], the Raman response comes from the intra-band scattering due to the near degeneracy of the lower and upper bands. Two peaks can be found: The high-energy one originates from Van Hove singularity \cite{lu2007} and the low-energy one from the SC gap opening along the Fermi surface. At intermediate temperature [Fig.~\ref{F6}(c)], both SC and DDW orders play significant roles. Apart from the intra-band contribution, the inter-band contribution, which is dominated by DDW, develops gradually. At high enough temperature where DDW dominates [Fig.~\ref{F6}(d)], the inter-band contribution takes over and the intra-band contribution is invisible. In Fig.~\ref{F6}(e), we combine the information about the peak positions and the peak intensities as a function of temperature. It is clear that the temperature evolution of Raman response exhibits a two-step pattern with an anomalous enhancement near the transition from the SC-dominated region to the DDW-dominated region.

To complete, in the heavily overdoped region the Raman peak energy follows the BCS prediction and decreases to zero as $T$ approaches $T_c$. The above results qualitatively agree with the experimental data on HgBa$_2$CuO$_{4+\delta}$ (Hg-1201) \cite{guyard2008-1,guyard2008,blanc2010}, as summarized in Fig.~\ref{F6}(f).

Most importantly, we found that the anomalous temperature enhancement of the peak energy in the B$_{1g}$ Raman response as $T \to T_c$ near $x^{}_{\text{QCP}}$ is intimately related to the back-bending of $T_d$ below the $T_{c}$ dome. It is nearly invisible for weak back-bending of $T_d$ and disappears in the original phase diagram. This may suggest the possible existence of the revised phase diagram in Hg-1201 where the anomalous temperature dependence of ESR peak energy is detected.

\section{Spin-density wave as a competing order}
\begin{figure}[tbp]
%\vspace{-0.0in} \hspace{-0.0in}
\center
\includegraphics[width=0.9\columnwidth,clip=true,angle=0]{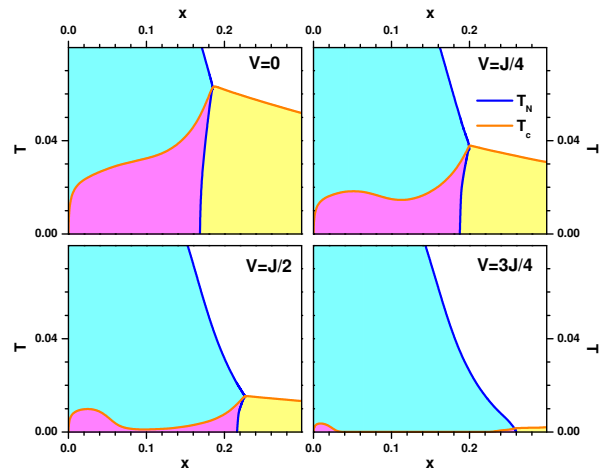}
%\vspace{-0.0in}
\caption{The $V$ dependence of the phase diagram in the color codes of yellow (SC), cyan (SDW), pink (coexisting) for $t^{\prime}=-0.25$, $t^{\prime\prime}=0.1$ and $J=0.35$.}
\label{SDW-V}
\end{figure}

We have presented the results for the CO being DDW, which has the $d$-wave symmetry. We also considered the competition between the SC and an $s$-wave-like order such as SDW.
Unlike DDW, the SDW order can be stabilized at low doping for $V=0$. Increasing $V$ will once again suppress the SC dome according to Eq.~\ref{V}, as shown in Fig.~\ref{SDW-V}.

\begin{figure}[tbp]
%\vspace{-0.0in} \hspace{-0.0in}
\center
\includegraphics[width=\columnwidth,clip=true,angle=0]{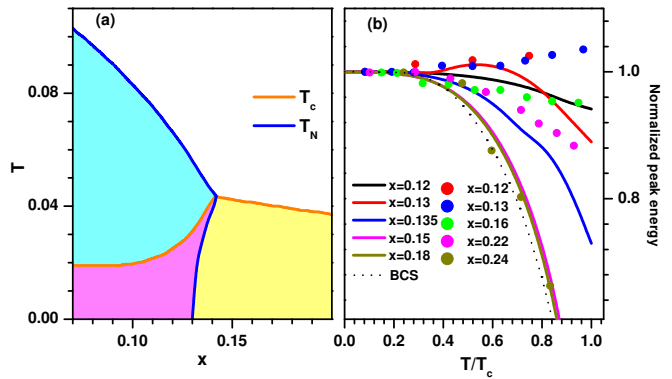}
%\vspace{-0.0in}
\caption{(a) Phase diagram and (b) Raman response in the extended $t$-$J$-$V$ model with an $s$-wave-like pseudogap SDW instead of the $d$-wave-like DDW. Symbols in (b) are experimental data extracted from cuprates Hg1201\cite{guyard2008-1,guyard2008}. $t^{\prime}=-0.25$, $t^{\prime\prime}=0.1$, $J=0.25$, and $V=0$.}
\label{F7}
\end{figure}

Fig.~\ref{F7}(a) presents a revised phase diagram that looks similar to the case of DDW as a CO. However, the SDW case exhibits considerably weakened $T_d$ back-bending under the $T_{c}$ dome in the intermediate doping range. The back-bending even disappears for certain parameters, giving rise to an original phase diagram. Meanwhile, the anomalous thermal evolution in the measured antinodal gap and in B$_{1g}$ Raman channel is also suppressed (Fig.~\ref{F7}(b)), consisting with the results found in DDW case. This may be understood as the case that DDW competes with SC more fiercely than SDW in the antinodal region. Hence, the pseudogap in the hole-doped cuprates is more likely to be a manifestation of DDW than SDW based on the mean-field theory, although it should be attested by rigorous numerical techniques.

\section{Discussion}
In Landau theory, the revised and original phase diagrams in high-$T_{c}$ superconductors can be established with the moderate and weak competitions, respectively. Thus, the question turns out to be whether the revised phase diagram does take place in real materials. In the basic $t$-$J$ model for cuprate superconductors, the pairing gap increases as the doping level decreases, promoting the notions of the pseudogap as a manifestation of preformed pairs and the $T_c$ dome as a manifestation of superconducting phase decoherence at low doping \cite{Anderson1987,zhang1988,Kotliar1988,Emery1995}. Inclusion of the nearest-neighbor Coulomb interaction $V$ favors DDW as the pseudogap state against SC in the underdoped region, leading to the formation of the $T_c$ dome structure in the phase diagram \cite{chakravarty2001,ubbens1992,yang2006,greco2009,cappelluti1999,bejas2012}. There are three possible sources for considerable $V$: (i) Strongly correlated metals are generally bad metals with large resistivity of the order of m$\Omega\cdot$cm and small optical Drude peak. Therefore, the electrostatic screening does not work well in those systems \cite{emery96,Varma1987,lawler10,frandsen14,yin16:FeTe}. (ii) In mean-field theory, the local constraint of no-double occupancy at each site is reinforced only globally. As a result, the expectation value of $\langle n_in_j \rangle $ is substantially greater than one for the undoped case. In this sense, $V$ acts to minimize this side effect of mean-field theory. (iii) More interestingly, upon mapping multiorbital real materials into a one-band effective low-energy Hamiltonian, a vacuum-fluctuation-induced effective interaction in the exactly same form as $V$ appears together with $J$ \cite{yin09,yin10:volja}. Like the superexchange $J$ term, the new `super-repulsion' $V$ term comes from virtual electron-hopping processes, which can hardly be screened electrostatically. The strength of super-repulsion $V$ is strongly material dependent, since the apical atoms are involved in the intermediate state of the vacuum charge fluctuation: $V/t$ was estimated to be $0.28$, $0.12$, and $0.08$ for apical oxygen (in La$_2$CuO$_4$),  chlorine (in Sr$_2$CuO$_2$Cl$_2$), and fluorine (in Sr$_2$CuO$_2$F$_2$), respectively \cite{yin09}. Our present calculations using this range of $V$ yield a revised phase diagram and electronic spectra consistent with ARPES and ERS measurements, indeed. Moreover, the realistic value of $t^{\prime}\sim -0.3$ is found to remove the pre-back-bending of $T_d$. Thus, it is necessary to include $V$ and $t^{\prime}$ in addressing the phase diagram of the real cuprate materials. The strong material dependence of $V$ renders the stability of DDW to be a material specific issue.

Following the above argument, $V$ in terms of effective low-energy Hamiltonian should be considerably strong in correlated electron systems in general. Like in the cuprates, $V$ may promote charge instabilities in the iron-based superconductors \cite{yin16:FeTe} in competition with SC. We notice that similar anomalous temperature dependence of ERS in the B$_{2g}$ channel was discovered in slightly underdoped Ba-122 iron-based superconductor \cite{chauviere2010} [open circles in Fig.~\ref{F6}(f)]. Together with the similar phase diagrams [Fig.~\ref{F1}(c)-(d)], this suggests the existence of strong competition between superconductivity and competing orders in iron-pnictide high-$T_{c}$ superconductors. Although the cuprates and iron pnictides appear very different from each other, e.g., in the properties of their parent materials, Fermi surface topology, forms of interactions, etc., they both exhibit strong phase competition. In fact, the active orbital physics in iron pnictides make the C$_{2}$ and C$_{4}$ competition more apparent in K-doped BaFe$_{2}$AS$_{2}$ or Na-doped SrFe$_{2}$AS$_{2}$ \cite{dai2012,taddei2016,allred2016,bohmer2015,avci2014}.

It is noteworthy that the present work has focused on the competition between SC and DDW/SDW. DDW was shown to be the leading possible charge instability in the one-band $t$-$t^{\prime}$-$J$-$V$ model \cite{bejas2012}. The recent Hall effect measurements on YBa$_2$Cu$_3$O$_y$ conducted at strong magnetic fields up to 88 tesla to suppress SC suggest that the pseudogap phase is disconnected from the charge-density wave (CDW) observed in the underdoped regime but linked to the antiferromagnetic Mott insulator \cite{Badoux2016}. This is not inconsistent with the DDW scenario, as DDW is not an ordinary CDW state whose order parameter is proportional to $\langle c^\dagger_i c_i \rangle$ or real $\langle c^\dagger_i c_j \rangle $ driven by Fermi surface instability, but a flux or bond-charge-phase order in terms of complex $\langle c^\dagger_i c_j \rangle $ due to the Mottness. For some other well-known COs such as loop-current order \cite{Varma1997} and intra-unit-cell nematic orders \cite{Fischer2011}, the three-band Emery model is an appropriate starting point. And it is yet to be seen whether the competition between SC and any other CO can produce a revised phase diagram and electronic spectra consistent with ARPES and ERS measurements in a realistic microscopic model.

\section{Summary}
We have shown in Ginzburg-Landau theory that the revised and original phase diagrams in high-$T_{c}$ superconductors can be established with the moderate and weak phase competitions, respectively. We further show that the revised phase diagram can result from the competition between DDW and SC or between SDW and SC in mean-field theory of the realistic $t$-$t^{\prime}$-$t^{\prime\prime}$-$J$-$V$ model. Inclusion of the much neglected feedback effect of SC on pseudogap can push the back-bending point from optimal doping to the overdoped regime. The calculated ARPES and ERS spectral functions reveal that the back-bending of $T^*$ can give a simple explanation of the observed anomalous temperature dependence of the antinodal gap via a two-step evolution where the SC and DDW dominate low- and high-temperature regions, respectively. Our results imply that it is likely to realize the revised phase diagram in cuprate superconductors.

\section{Acknowledgments}
We thank Peter D. Johnson, J.-X. Li, and Z.-X. Shen for helpful discussions and suggestions. This work was supported by the National Nature Science Foundation of China under Contract No. 11274276, the Ministry of Science and Technology of China 2016YFA0300401, and the U.S. Department of Energy (DOE), Office of Basic Energy Science, under Contract No. DE-SC0012704. Y. Zhou acknowledges the financial support of CSC and visiting scholarship of Brookhaven National Laboratory. H.Q. Lin acknowledges support from NSAF U1530401 and computational resource from the Beijing Computational Science Research Center.

\section*{Appendix: Solving the extended $t$-$J$-$V$ model}
The extended $t$-$J$-$V$ model is solved in mean-field-type theories with the order parameters defined as follows: (i) The $d$-wave SC order $\frac{1}{2}\langle c_{i\uparrow} c_{j\downarrow}-c_{i\downarrow} c_{j\uparrow}\rangle=\pm \Delta$ with $+$ for the $x$-direction and $-$ for the $y$ direction, (ii) the uniform bond order and the DDW order $\langle c_{i}^\dagger c_{j}\rangle=\chi\pm iD$ with $+$ for the $x$ direction of the $A$ sublattice and the $y$-direction of the $B$ sublattice, and $-$ otherwise (see Fig.~\ref{A1}), and (iii) the SDW order $\frac{1}{2}\langle c_{i\uparrow}^\dagger c_{i\uparrow}-c_{i\downarrow}^\dagger c_{i\downarrow}\rangle=(-1)^{i} m$. The interacting terms $H_{JV}=J\sum_{\langle i,j\rangle}\Big(\vec{S}_i \cdot \vec{S}_j-\frac{1}{4}n_in_j\Big)+V\sum_{\langle i,j\rangle}n_{i} n_{j}$ are decoupled into the particle-particle and particle-hole channels \cite{ubbens1992}:
\begin{eqnarray}
H_{JV}&=&-V_{c}\sum_{\langle i,j\rangle}[\Delta(c_{i\downarrow}^{\dagger}c_{j\uparrow}^{\dagger}-c_{j\downarrow}^{\dagger} c_{i\uparrow}^{\dagger})+h.c.]\nonumber\\
&-&V_{d}\sum_{\langle i,j\rangle}[(\chi\pm iD)(c_{j\uparrow}^{\dagger}c_{i\uparrow}+c_{j\downarrow}^{\dagger}c_{i\downarrow})+h.c.]\nonumber\\
&+&2Jm\sum_{i}{(-1)^{i}(c_{i\uparrow}^{\dagger}c_{i\uparrow}-c_{i\downarrow}^{\dagger}c_{i\downarrow})}\text{,}
\end{eqnarray}
where $V_{c}=J-V$ and $V_{d}=J/2+V$.
	
\subsection{Slave-boson approximation}
In the slave-boson approximation, the physical electron operators $c_{i\sigma}=b_{i}^{\dagger}f_{i\sigma}$ are represented by slave bosons $b_{i}$ carrying the charge and fermions $f_{i\sigma}$ representing the spin $\sigma$ with the constraint $\sum_{\sigma}f_{i\sigma}^{\dagger}f_{i\sigma}+b_{i}^{\dagger}b_{i}=1$\cite{Brinckmann1999}. In mean-field theory, bosons condense $b_{i} \rightarrow \langle b_{i} \rangle=\sqrt{x} $ with $x$ the hole concentration. The mean-field Hamiltonian is then expressed in momentum space as
\begin{eqnarray}
H=\sum_{k}\psi _{k}^{\dagger }\left(
\begin{array}{cccc}
\varepsilon _{k} & -iD_{k} & \Delta _{k} & 0 \\
iD_{k} & \varepsilon _{k+Q} & 0 & -\Delta _{k} \\
\Delta _{k} & 0 & -\varepsilon _{k} & -iD_{k} \\
0 & -\Delta _{k} & iD_{k} & -\varepsilon _{k+Q}%
\end{array}%
\right) \psi _{k}\text{,}
\label{SE4}
\end{eqnarray}
where $\psi _{k}=\left( f_{k\uparrow }\text{ }f_{k+Q\uparrow }\text{ }%
f_{-k\downarrow }^{\dagger }\text{ }f_{-k-Q\downarrow }^{\dagger }\right)
^{T}$ with $Q=\left( \pi \text{, }\pi \right) $ being the antiferromagnetic wave
vector. $\varepsilon _{k}=-2\left( xt+V_{d}\chi \right) \left( \cos k_{x}+\cos
k_{y}\right) -4xt^{\prime }\cos k_{x}\cos k_{y}-2xt\left( \cos 2k_{x}+\cos
2k_{y}\right) -\mu $ with $t$, $t^{\prime }$, and $t^{\prime \prime }$ being the
nearest-, next-nearest-, and third-nearest-neighbor hopping constants, respectively. $D_{k}=2V_{d}D(%
\cos k_{x}-\cos k_{y})$, and $\Delta _{k}=2V_{c}\Delta (\cos k_{x}-\cos k_{y})$. The
summation is restricted in the magnetic Brillouin zone.

The order parameters can be self-consistently determined by minimizing the free energy
\begin{eqnarray}
F&=&-\frac{2T}{N}\sum_{k,\eta =\pm }^{\prime }\ln (2\cosh \frac{\beta
E_{k}^{\eta }}{2})-\mu(1-x) \nonumber\\
&+&(4V_{d}\chi ^{2}+4V_{d}D^{2}+4V_{c}\Delta ^{2})\text{.}
\end{eqnarray}
Here $E_{k}^{\eta }=\sqrt{\left( \xi _{k}^{\eta }\right) ^{2}+\Delta _{k}^{2}}$ with $\xi _{k}^{\eta }=\left( \frac{%
\varepsilon _{k}+\varepsilon _{k+Q}}{2}\right) +\eta \sqrt{\left( \frac{%
\varepsilon _{k}-\varepsilon _{k+Q}}{2}\right) ^{2}+|D_{k}|^{2}}$ ($\eta =1$ and $-1$ for upper and lower band, respectively) is obtained by unitary transformation with the $4 \times 4$ matrix $U_{k}$ \cite{yuan2006}.

When the incommensurate DDW order is included, the phase boundary is determined by the charge order instability under the random phase approximation (RPA). The RPA charge susceptibility for DDW is
\begin{equation}
	\chi_{RPA}(i\nu,q)=\frac{\chi_{0}(i\nu,q)}{1-(\frac{J}{4}+\frac{V}{2})\chi_{0}(i\nu,q)}.
\end{equation}
Here, the bare charge susceptibility for DDW is $\chi_{0}(\tau,q)=\langle T \rho_{q}(\tau)\rho_{q}^{\dagger}(0) \rangle_{0}$ with $\rho_{q}(\tau)=\sum_{k\sigma}i(\sin(k_{x}-\frac{q_{x}}{2})-\sin(k_{y}-\frac{q_{y}}{2}))f_{k+q\sigma}^{\dagger}(\tau)f_{k\sigma}(\tau)$. The charge instability is therefore judged from the divergency of the RPA charge susceptibility at zero frequency, yielding the simple criterion of $\mathcal{D}(q)=1-(\frac{J}{4}+\frac{V}{2})\chi_{0}(0,q)=0$ with $\mathcal{D}(q)$ the denominator at zero frequency. Here, $q=(\pi,\pi-\delta q)$ with $\delta q=0$, and $\delta q\ne 0$ for the commensurate, and incommensurate DDW, respectively.

\begin{figure}[tbp]
%\vspace{-0.0in} \hspace{-0.0in}
\center
\includegraphics[width=\columnwidth,clip=true,angle=0]{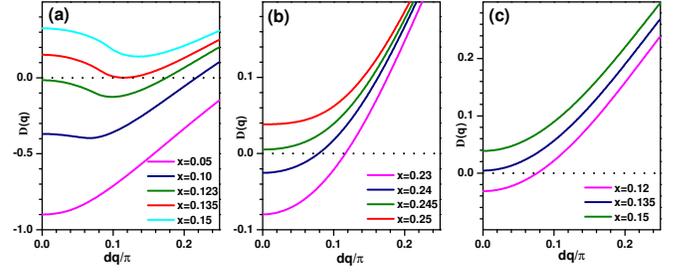}
%\vspace{-0.0in}
\caption{(a) The denominator of RPA DDW charge susceptibility $\mathcal{D}(q)$ with different parameters. (a) $t^{\prime}=0$, $t^{\prime\prime}=0$, $J=0.3$ and $V=0.15$; (b) $t^{\prime}=-0.35$, $t^{\prime\prime}=0$, $J=0.3$ and $V=0.15$; and (c) $t^{\prime}=-0.25$, $t^{\prime\prime}=0.1$, $J=0.3$ and $V=0.12$. The temperature is fixed at $1\times 10^{-4}$.}
\label{SF}
\end{figure}

Fig.~\ref{SF} shows some relevant results at low enough temperature. The data containing only the nearest neighbor hopping is shown in Fig.~\ref{SF}(a). The instability of the commensurate DDW order occurs at doping density about $x=0.123$ where $\mathcal{D}(q)=0$ with $dq=0$. In comparison, the incommensurate DDW instability occurs at about $x=0.135$ with $dq\sim0.1\pi$ (also see Fig.~\ref{F45}(c) in main text). This is well consistent with the previous results obtained by large-N expansion in absence of superconductivity\cite{bejas2012}, manifesting the existence of the incommensurate DDW order. It had been shown that the incommensurate DDW order is strongly weakened by introducing the next-nearest neighbor hopping\cite{bejas2012} (Fig.~\ref{F45}(b) in main text). Furthermore, the incommensurate DDW order may be further suppressed by SC as shown in Fig.~\ref{SF}(b) (Also Fig.~\ref{F45}(d)), no instability of the incommensurate DDW state is found when the superconductivity is included. The DDW instability in the SC state with the parameters presented in the main text is shown in Fig.~\ref{SF}(c), only the commensurate DDW instability occurs at about $x=0.135$.

The spectral function $A(k,\omega ) =-\frac{1}{\pi }\Im G^{11}(k,i\omega _{l}\rightarrow \omega
+i\Gamma )$ is calculated with the Matsubara Green function
\begin{equation}
G^{nm}(k,i\omega _{l})=\sum_{j=1}^{4}\left( U_{k}\right) _{nj}\frac{1}{%
i\omega _{l}-E_{k}^{j}}\left( U_{k}^{\dagger }\right) _{jm}
\end{equation}

The Raman response is described by the
following Matsubara correlation function \cite{lu2007}
\begin{equation}
R_{\gamma }(q,\tau )=-\left\langle \mathrm{T}\rho _{\gamma _{k}}(q,\tau
)\rho _{\gamma _{k}}(-q,0)\right\rangle\text{,}
\end{equation}%
where $\rho _{\gamma }(q,\tau )=\sum_{k}f_{k+q/2}^{\dagger }(\tau )\gamma
_{k}f_{k-q/2}(\tau )$ with the vertex $\gamma _{k}=\frac{1}{2}(\frac{\partial
^{2}\varepsilon _{k}}{k_{x}^{2}}-\frac{\partial ^{2}\varepsilon _{k}}{%
\partial k_{y}^{2}})$ for the B$_{1g}$ channel and $\gamma _{k}=\frac{\partial
^{2}\varepsilon _{k}}{\partial k_{x}\partial k_{y}}$ for the B$_{2g}$ channel. At the zero-momentum transfer $R_{\gamma }(0,i\omega _{l})$
corresponds to what ERS experiments measure
\begin{equation}
R_{\gamma }(0,i\omega _{l})=\sum_{k,n,m}\frac{f(E_{k}^{n})-f(E_{k}^{m})}{%
i\omega _{l}+E_{k}^{m}-E_{k}^{n}}\vert(\text{\textrm{U}}_{k}^{\dagger }%
\boldsymbol{\gamma }_{k}\mathrm{U}_{k})_{nm}\vert^{2}
\end{equation}
with $f(E_{k}^{n})$ the Fermi-Dirac function.

\subsection{Renormalized mean-field theory}
The renormalized mean-field theory (RMFT) projects the Hamiltonian by Gutzwiller factors. The expectation value of the projected Hamiltonian is
\begin{eqnarray}
\langle H \rangle&=&-\sum_{ij\sigma}g^{t}t_{ij} \langle \chi +iD\rangle -\mu\sum_{i\sigma}\langle n_{\sigma}\rangle\nonumber\\
		&-&2[(\frac{1}{2}g^{xy}+\frac{1}{4}g^{z})J-V]\sum_{\langle i,j\rangle}\langle \Delta\rangle\langle \Delta\rangle^{*}\nonumber\\
		&-&2[(\frac{1}{2}g^{xy}+\frac{1}{4}g^{z})J+V]\sum_{\langle i,j\rangle}(\chi+iD)^{*}(\chi+iD)\nonumber\\
		&+&g^{z}J\sum_{\langle i,j\rangle}\langle m_{i}\rangle \langle m_{j}\rangle\text{,}
\end{eqnarray}
where $\chi$, $D$, $\Delta$ and $m$ are variational parameters (their sign rules are the same as those specified in the last subsection). $g^{t}$, $g^{xy}$, $g^{z}$ are the Gutzwiller factors for hopping, transverse and longitudinal spin-exchange terms, respectively. The expectation value of an operator $O$ in the projected state is $g^{O}\langle O \rangle$ with $\langle O \rangle$ is the expectation value in the unprojected state and $g^{O}$ is the Gutzwiller factor for operator $O$. Here, $g^{\Delta}=(g^{t})^{2}$, $g^{m}=\sqrt{g^{z}}$, $g^{\chi,D}=g^{t}$.

In fact, the simplest Gutzwiller approximation \cite{zhang1988} does not reproduce the results obtained by variational Monter Carlo method. For example, the resulting antiferromagnetic state extends to high doping density. It can be improved by taking the feedback effect of the order parameters into account \cite{ogata2003,yang2009}. The modified Gutzwiller factors are
\begin{eqnarray}
	&g^{t}(i,j)=\frac{2x}{1+x}\nonumber\\
	&g^{xy}=g^{z}=\left( \frac{2}{1+x} \right)^{2}a^{-7}\text{,}
\end{eqnarray}
where $a=1+\frac{4X}{(1-x^{2})^{2}}$ with $X=2x^{2}(\Delta^{2}-|\chi+iD|^{2})+2(|\chi+iD|^{2}+\Delta^{2})^{2}$.

At finite temperature, one should minimize the free energy $F=\langle H \rangle -TS$ instead. $\langle H \rangle$ is straightforward by using finite temperature Wicks theorem. $S=S_{0}+\delta S$ with $S_{0}=-\sum_{n}[f(E_{n})\ln f(E_{n})+(1-f(E_{n}))\ln (1-f(E_{n}))]$ is the entropy in the mean-field trial state, $f$ is the Fermi-Dirac distribution function, and $\delta S$ is the entropy which losses under projection as \cite{wang2010}
\begin{equation}
\delta S=-N\left( x\ln \frac{4x}{(1+x)^{2}+4m}+(1-x)\ln\frac{2(1-x)}{1-x^{2}+4m} \right).
\end{equation}
For the nonmagnetic case, $\delta S$ is temperature independent and thus can be ignored.

\bibliography{ref}

\end{document}